%% file: ISLPED26_WarPGNN.tex
\documentclass[sigconf,screen=true,bookmarks=false]{acmart}
\usepackage{algorithm}
\usepackage{pifont}

\input{packages}
\input{macro}

\graphicspath{{./figs/}}
\usepackage{geometry}
\usepackage{tikz}
\usepackage{xcolor}

\newcommand{\blackcircled}[1]{%
    \raisebox{0.2ex}{%
  \tikz[baseline=(char.base)]{
    \node[shape=circle, fill=black, inner sep=0.5pt] (char)
      {\color{white}\footnotesize #1};}%
}
}

\begin{document}

\settopmatter{printacmref=false} 
\renewcommand\footnotetextcopyrightpermission[1]{} 

\pagestyle{plain} 

\title{{WarPGNN}: A Parametric Thermal \underline{War}page Analysis Framework with \underline{P}hysics-aware \underline{G}raph \underline{N}eural \underline{N}etwork}
\author{Haotian Lu, Jincong Lu, Sachin Sachdeva and Sheldon X.-D. Tan}
\affiliation{%
  \institution{
  Department of Electrical and Computer Engineering
  \\University of California, Riverside}
  \city{Riverside, CA}
  \country{USA}
}
\email{hlu123@ucr.edu, stan@ece.ucr.edu}

\input{doc/0abstract}



\maketitle

\input{doc/1intro}

\input{doc/2prelim}
\input{doc/3method}

\input{doc/4result}

\input{doc/5conclu}

\balance
\bibliographystyle{unsrt}
\bibliography{ref/EM_basics, ref/mscad, ref/neural_network, ref/warpage}

\end{document}

%% file: packages.tex
\usepackage[utf8]{inputenc} 
\usepackage[T1]{fontenc}    
\usepackage{pifont}

\usepackage{algorithm}
\usepackage{algorithmicx}
\usepackage[noend]{algpseudocode}
\usepackage{graphicx}
\usepackage{threeparttable}
\usepackage{multirow}
\usepackage{amsmath}
\usepackage{bm}
\usepackage{bbm}
\usepackage{amsthm}
\usepackage{enumitem} 
\usepackage[subrefformat=parens,labelformat=parens]{subfig}

\usepackage{wrapfig}


%% file: macro.tex
\definecolor{citecolor}{RGB}{34,139,34}
\definecolor{mydarkblue}{rgb}{0,0.08,1}
\definecolor{mydarkgreen}{rgb}{0.02,0.6,0.02}
\definecolor{mydarkred}{rgb}{0.8,0.02,0.02}
\definecolor{mydarkorange}{rgb}{0.40,0.2,0.02}
\definecolor{mypurple}{RGB}{111,0,255}
\definecolor{myred}{rgb}{1.0,0.0,0.0}
\definecolor{mygold}{rgb}{0.75,0.6,0.12}
\definecolor{myblue}{rgb}{0,0.2,0.8}
\definecolor{mydarkgray}{rgb}{0.,0.2,0.2}

\definecolor{lightred}{RGB}{255,235,235}
\definecolor{lightgreen}{RGB}{235,255,235}
\definecolor{lightblue}{RGB}{235,235,255}
\definecolor{lightcyan}{RGB}{235,255,255}
\definecolor{lightmagenta}{RGB}{255,235,255}
\definecolor{lightyellow}{RGB}{255,255,235}

\definecolor{qxkcolor}{RGB}{215,235,255}
\definecolor{softmaxcolor}{RGB}{230,235,255}
\definecolor{probxvcolor}{RGB}{255,255,235}

\definecolor{topkcolor}{RGB}{255,235,235}
\definecolor{zecolor}{RGB}{255,255,235}
\definecolor{dynacolor}{RGB}{235,255,255}

\definecolor{reviewcolor}{RGB}{0,0,200}

\theoremstyle{plain}

\theoremstyle{definition}

\newcommand{\squishlist}{
 \begin{list}{$\bullet$}
  { \setlength{\itemsep}{0pt}
     \setlength{\parsep}{3pt}
     \setlength{\topsep}{3pt}
     \setlength{\partopsep}{0pt}
     \setlength{\leftmargin}{1.5em}
     \setlength{\labelwidth}{1em}
     \setlength{\labelsep}{0.5em} } }
     
\newcommand{\squishend}{
  \end{list}  }     



%% file: doc/0abstract.tex
\begin{abstract}
\label{abstract}

With the advent of system-in-package (SiP) chiplet-based design and heterogeneous 2.5D/3D integration, thermal-induced warpage has become a critical reliability concern. 
While conventional numerical approaches can deliver highly accurate results, they often incur prohibitively high computational costs, limiting their scalability for complex chiplet-package systems. 
In this paper, we present {\it \textbf{WarPGNN}}, an efficient and accurate parametric thermal warpage analysis framework powered by Graph Neural Networks (GNNs). 
By operating directly on graphs constructed from the floorplans, {\it \textbf{WarPGNN}} enables fast warpage-aware floorplan exploration and exhibits strong transferability across diverse package configurations.
Our method first encodes multi-die floorplans into reduced Transitive Closure Graphs (rTCGs), then a Graph Convolution Network (GCN)-based encoder extracts hierarchical structural features, followed by a U-Net inspired decoder that reconstructs warpage maps from graph feature embeddings.
Furthermore, to address the long-tailed pattern of warpage data distribution, we developed a physics-informed loss and revised a message-passing encoder based on Graph Isomorphic Network (GIN) that further enhance learning performance for extreme cases and expressiveness of graph embeddings.  
Numerical results show that \textbf{\textit{WarPGNN}} achieves more than 205.91$\times$ speedup compared with the 2-D efficient FEM-based method and over 119766.64$\times$ acceleration with 3-D FEM method COMSOL, respectively, while maintaining comparable accuracy at only 1.26\% full-scale normalized RMSE and 2.21\% warpage value error.
Compared with recent DeepONet-based model, our method achieved comparable prediction accuracy and inference speedup with 3.4$\times$ lower training time.
In addition, \textbf{\textit{WarPGNN}} demonstrates remarkable transferability on unseen datasets with up to 3.69\% normalized RMSE and similar runtime.
\end{abstract}

%% file: doc/1intro.tex
\vspace{-5pt}
\section{Introduction}
\label{sec:Introduction}

Due to the increasing demand for high-performance computing, 2.5D and 3D chiplet-based integration that packs various components fabricated with different technologies becomes a promising technique for beyond-Moore computing paradigm due to its modular designs, multiple functionalities, low cost and few manufacturing defects~\cite{Naffziger_ISSCC20}.
However, the integration of heterogeneous components and packages incurs emerging reliability concerns.
Among them, thermal warpage has become a more critical issue at the package level, where both homogeneous and heterogeneous integrations are prone to deformation caused by the Coefficient of Thermal Expansion (CTE) mismatch between various materials under thermal load~\cite{Lau_TCPMT18, Lau_TCPMT22}.
An illustration of the warpage formation is shown in Fig.~\ref{fig:thermal_warpage}.

An efficient while accurate thermal warpage analysis is critical to optimize the floorplan design of chiplet-based systems to improve reliability~\cite{Hsu_ICCAD22_TCGFloorplan, Zhuang_ICCAD22, Chen_TCAD24}.
Many analytical as well as numerical solvers have been proposed to simulate thermal warpage problem.
Suhir's 1D model and its revisions~\cite{Mishkevich_1993_Suhir, Tsai_JEP04, Tsai_TCPMT20} utilized physics-based closed-form analytical solution to calculate deformation value and curvature at specific positions with relatively high accuracy.
However, such analytical models are always over-simplified, being applicable only to one-dimensional analyses with fixed layer (two layers) and unsuitable for higher-dimensional cases with complex structures with multiple layers. 
Commercial solvers such as ANSYS~\cite{ANSYS2024} and COMSOL~\cite{comsol:heat'2014}, and other Finite Element Method (FEM) based solvers~\cite{Shen_tcmpt16, McCann_tcmpt17, Kanemoto_ECTC18, Hao_IPFA21} can accurately capture displacement for complex structures, but demand high computation overhead.
To further improve computational efficiency, Lo \textit{et al.}~\cite{Lo_ICCAD24_EfficientWarpage} proposed a more efficient FEM-based solver that first transforms three-dimensional structure to a 2-D plane, and solves it based on 2-D FEM formulation leveraging  an efficient Discrete Kirchhoff Triangle (DKT) theory~\cite{Bhothikhun03072015} for flip-chip package based designs. This method still suffers from FEM-related high computational cost. 
Recently,~\cite{Lo_ICCAD25_DeepONet} addressed this challenge using a DeepONet~\cite{Lu2019LearningNO}-based surrogate for advanced-package warpage prediction, demonstrating that learning-based approaches can significantly reduce FEM cost while preserving high accuracy, albeit with heavy training overhead.
\input{figtex/warpage_illu}

On the other hand, recent breakthroughs in deep learning for cognitive tasks based on deep neural networks (DNNs)~\cite{LeCun:2015dt} have opened up new opportunities for various applications in electronic design automation (EDA)~\cite{Yu_MLCAD23, MaskPlace, Lu:ISLPED25, Lamichhane:ISQED26}.
Among them, Graph Neural Networks (GNNs) have emerged as a key architecture that leverages graph representations to perform various learning tasks~\cite{Lu_DAC20, Ren_DAC20, Jin:DAC'21, Chen:ASPDAC'22, zuo2024graph, guo2025irgnn}.
Specifically, \textit{EMGraph}~\cite{Jin:DAC'21} encoded multi-segment interconnect structures into graph representations and trained a GNN based on GraphSAGE~\cite{Hamilton_NIPS17_GraphSAGE} to predict stress distributions, with order-of-magnitude of speedup and marginal accuracy loss.
\textit{ThermGCN}~\cite{Chen:ASPDAC'22} based on Graph Convolution Network (GCN)~\cite{kipf2017semi} is proposed to predict full-chip thermal maps for chiplet-based 2.5D systems, achieving remarkable speedup compared to conventional numerical methods and demonstrated strong knowledge transferability.
Guo~\textit{et al.}~\cite{guo2025irgnn} proposed a graph-based \textit{IRGNN} that integrates a numerical solver and point clouds while exploiting power-grid topology for efficient IR-drop analysis.
These studies collectively demonstrate the exceptional capability and generalizability of GNNs in performing high-fidelity graph-to-field prediction.

In this work, to mitigate the high computing costs in FEM while preserving sufficient accuracy in thermal warpage simulation, we propose a data-driven physics-aware parametric framework termed \textbf{\textit{WarPGNN}} for efficient thermal warpage prediction of flip-chip packages, leveraging GNN's inherent capability in processing graph data and strong transferability.
We employ the widely-accepted encoder-decoder architecture~\cite{Hinton2006Autoencoder} with a GCN-based encoder for graph feature embedding and a decoder inspired by U-Net~\cite{ronneberger2015unet} for deformation reconstruction.
Our major contributions in this work are as follows:
\begin{itemize}[leftmargin=1em, labelsep=0.5em, parsep=2.5pt]

\item We propose the first GNN-based parametric framework to perform efficient and accurate thermal warpage analysis for multi-die floorplans, leveraging a lightweight reduced Transitive Closure Graph (rTCG) format to represent multi-die placement as shown in Fig.~\ref{fig:floorplan}. 
The resulting \textbf{\textit{WarPGNN}} enables fast parametric design-space exploration for warpage-aware floorplan optimization.

\item We formulate the thermal warpage simulation problem as a graph-to-field prediction task.
Our backbone model contains an encoder based on MLP-structured GCN that takes rTCG as input, and a decoder with U-Net architecture is cascaded to remap node embeddings to out-of-plane deformation map.

\item We further investigate the Partial Differential Equation (PDE)-governed characteristics of warpage data, and design a physics-informed loss that accounts for the long-tailed data distribution and field smoothness for accuracy enhancement, especially on extreme cases.
Additionally, a GIN-based message-passing encoder with higher expressive capability for graph embeddings is developed to further enhance the overall prediction accuracy.

\item 
Numerical results show that \textbf{\textit{WarPGNN}} achieves a computational speedup of over 119766.64$\times$ compared to FEM-based COMSOL~\cite{comsol:heat'2014}, and a 205.91$\times$ acceleration over the efficient 2-D FEM solver~\cite{Lo_ICCAD24_EfficientWarpage}, with only 1.26\% full-scale normalized RMSE and 2.21\% warpage error compared to COMSOL.
Additionally, \textbf{\textit{WarPGNN}} reaches similar performance with 3.4$\times$ lower training time compared with DeepONet-based method~\cite{Lo_ICCAD25_DeepONet}.
Furthermore, our method demonstrates strong transferability on four unseen datasets, achieving normalized RMSE of up to 3.69\% in deformation map prediction.
\end{itemize}

%% file: figtex/warpage_illu.tex

\begin{figure}[tp]
    \centering
    \vspace{-15pt}
    \subfloat[]{
        \centering
        \includegraphics[width=0.5\columnwidth]{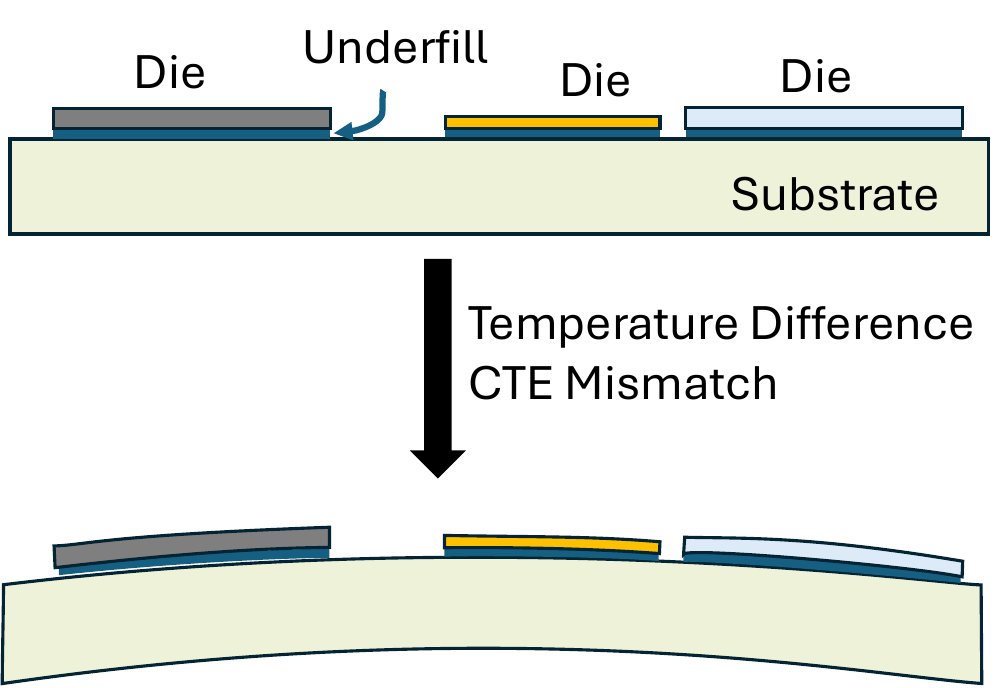}
        \label{fig:thermal_warpage}
    }\hspace{3pt}
    \subfloat[]{
        \centering
        \includegraphics[width=0.35\columnwidth]{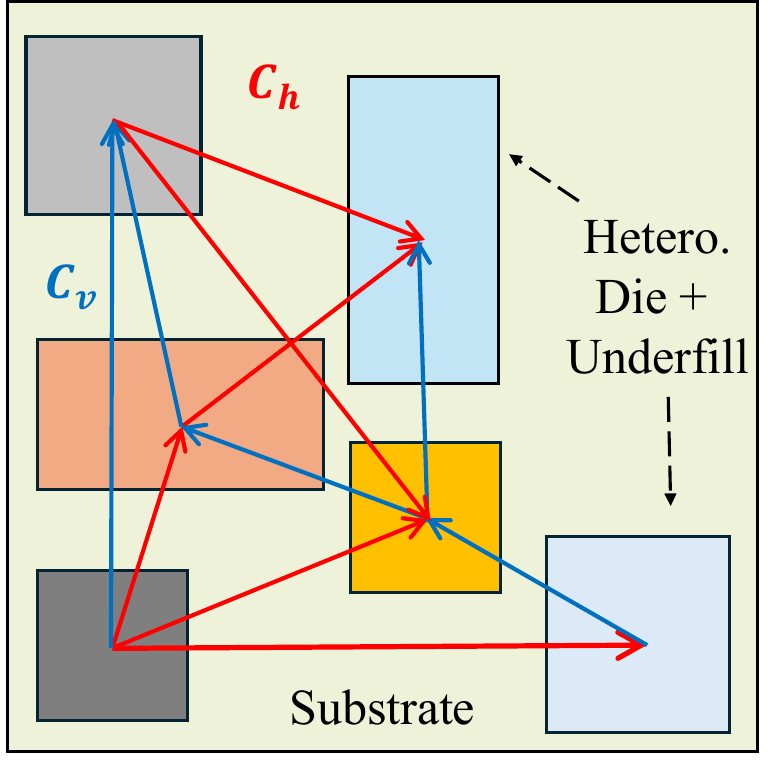}
        \label{fig:floorplan}
    }
    \vspace{-10pt}
    \caption{\small (a) Illustration of thermal warpage formation. 
    (b) TCG representation for a 6-die floorplan.
    Closure edges have been eliminated.}
    \vspace{-20pt}
\end{figure}

%% file: doc/2prelim.tex
\section{Preliminaries}
\label{sec:prelim}
\subsection{Thermal Warpage Problem for Package Design}
The thermal warpage in a typical flip-chip package of die-underfill-substrate assembly considered in this work is defined as the deformation from its original shape.
Package warpage refers to the value difference between the maximum and the minimum deformations along the z-axis, shown below in Eq.~\eqref{eq:warpage_def}~\cite{Lo_ICCAD24_EfficientWarpage},
\begin{equation}
\small
    W=\max_{(x,y)\in\Omega} w(x,y) - \min_{(x,y)\in\Omega} w(x,y)
    \label{eq:warpage_def}
\end{equation}
here $w(x,y)$ marks out-of-plane deformation of the z-axis, $x$ and $y$ represent the geometrical coordinates on deformation surface and $\Omega$ denotes the domain of package.

\subsection{3-D FEM Numerical Analysis Method}
For warpage analysis, several simplified analytical models based on one-dimensional beam theory have been proposed~\cite{Tsai_TCPMT20}.
However, as package architectures become more heterogeneous and multi-physics interactions intensify, these analytical formulations exhibit increasing discrepancies compared to high-fidelity thermo-mechanical simulations. Consequently, FEM solvers remain the predominant approaches for accurate warpage prediction, albeit at the expense of substantial computational cost.
For 3-D warpage analysis, the FEM formulation is shown below~\cite{Bathe2006}
\begin{equation}
\small
    [K]\{d\} = \{F_{\mathrm{th}}\}
\label{eq:warpage_3D_FEM}
\end{equation}
where $d$ is the displacement, and the stiffness matrix $[K]$ and thermal load vector $F_{\mathrm{th}}$ are assembled from the element-level stiffness matrix and thermal load vector defined in Eq.~\eqref{eq:thermal_load},
\begin{align}
\small
    [K_e] &= 
    \int_{\Omega_e} [B]^T [C] [B]\, d\Omega; \,\,\,
    \{F_e^{\mathrm{th}}\} = 
    \int_{\Omega_e} [B]^T [C] \, \{\varepsilon^{\mathrm{th}}\} \, d\Omega.
\label{eq:thermal_load}
\end{align}
here $\varepsilon^{\mathrm{th}}$ is the strain due to thermal load. 
The matrix $C$ represents the strain to stress relation due to the mismatched  coefficient of thermal expansion (CTE) of adjacent layers at the element level. 
Matrix $[K]$ in Eq.~\eqref{eq:warpage_3D_FEM} typically is singular due to rigid-body issue. 
Rigid-body motion is prevented by minimal constraints
(one node fixed in all directions, two additional nodes fixed selectively).
After solving this equation, the warpage is just the displacement in $z$ direction.  
\subsection{2-D FEM Analysis based on Laminate Theory}
To achieve a better trade-off between efficiency and analysis accuracy,
~\cite{Lo_ICCAD24_EfficientWarpage} employs Classical Laminate Theory (CLT)~\cite{reddy2003mechanics} for multi-layer package warpage analysis as thickness of package layers are much smaller than package dimension~\cite{Huang_ECTC21}. The idea is to integrate the thickness and forming "resultants" including forces and their moments in the formulation. The resulting energy minimization equation is shown in Eq.~\eqref{eq:6a},
\begin{equation}
\small
\Pi =
\iint_{\Omega}
\frac{1}{2}
\begin{pmatrix}
\boldsymbol{\varepsilon}\\
\mathbf{k}
\end{pmatrix}^{\!T}
\begin{pmatrix}
\mathbf{A} & \mathbf{B}\\
\mathbf{B} & \mathbf{D}
\end{pmatrix}
\begin{pmatrix}
\boldsymbol{\varepsilon}\\
\mathbf{k}
\end{pmatrix}
-
\begin{pmatrix}
\mathbf{N}_t\\
\mathbf{M}_t
\end{pmatrix}^{\!T}
\begin{pmatrix}
\boldsymbol{\varepsilon}\\
\mathbf{k}
\end{pmatrix}
\,dx\,dy,
\label{eq:6a}
\end{equation}
where the in-plane strain ${\varepsilon}(x,y)$ and  curvature vectors $\mathbf{k}(x,y)$ are defined as Eq.~\eqref{eq:6b},
\begin{equation}
\small
\boldsymbol{\varepsilon}(x,y)=
\begin{bmatrix}
\frac{\partial u(x,y)}{\partial x}\\[4pt]
\frac{\partial v(x,y)}{\partial y}\\[4pt]
\frac{\partial u(x,y)}{\partial y} + \frac{\partial v(x,y)}{\partial x}
\end{bmatrix}, \quad
\mathbf{k}(x,y)=
\begin{bmatrix}
-\frac{\partial^2 w(x,y)}{\partial x^2}\\[4pt]
-\frac{\partial^2 w(x,y)}{\partial y^2}\\[4pt]
-2\,\frac{\partial^2 w(x,y)}{\partial x\,\partial y}
\end{bmatrix},
\label{eq:6b}
\end{equation}
\noindent
where $u(x,y)$, $v(x,y)$, and $w(x,y)$ denote the displacements of the laminate reference plane in the $x$, $y$, and $z$ directions, respectively.  $A$, $B$ and $D$ are the extensional, coupling, and bending stiffness matrices
served as a connection between the applied thermal loads and the strains
in the laminate. By assuming in-plane strain ${\varepsilon}(x,y) = 0$~\cite{WU2018582}, 
the governing equation can be derived by minimizing revised energy function shown in Eq.~\eqref{eq:DKT},
\begin{equation}
\small
\Pi^* = 
\iint_{\Omega} 
\left(
\tfrac{1}{2}\mathbf{k(x,y)}^T D^*(x,y)\,\mathbf{k(x,y)}
- \mathbf{M}_t^*(x,y)^T\mathbf{k(x,y)}
\right)
\,dx\,dy,
\label{eq:DKT}
\end{equation}
where $D^*(x,y)$ denotes the reduced bending stiffness matrix obtained from the CLT parameters, and $\mathbf{M}_t^*(x,y)$ represents the effective thermal moment (loads). 

The 2-D FEM indeed leads to significant speedup over the 3-D FEM method; however, this method still requires mesh generation and matrix solving for each new design, as it is still a conventional FEM solver suffering from computational inefficiency.

%% file: doc/3method.tex
\section{Proposed \textit{WarPGNN} for Parametric Thermal Warpage Analysis}
\label{sec:model}

\subsection{Graph Encoding for Multi-die Placement}
We formulate the thermal warpage analysis problem as graph-to-field regression task, as multi-die floorplans with various die assignments can be naturally viewed as graphs.
In this work, we adopt the Transitive Closure Graph (TCG) representation for graph encoding, which has been widely used to represent multi-die flip-chip floorplans~\cite{Lin_DAC01, Lin_tcad02_TCG-S, Hsu_ICCAD22_TCGFloorplan}.
TCG consists of an acyclic horizontal TCG $C_h$ and an acyclic vertical TCG $C_v$, modeling the geometric relations between all dies as shown in Fig.~\ref{fig:floorplan}.

To improve the computational efficiency during GNN training, we employ a reduced Transitive Closure Graph (rTCG) generation flow as shown in Algo.~\ref{alg:rTCG_short} adapted from~\cite{Lin_TVLSI04_TCG}, which removes redundant edges while strictly preserving all geometric adjacency relations among dies. 
The procedure first invokes $\textsc{InsertPrecedeEdge}$ to examine each die pair $(i,j)$ and determine their horizontal or vertical precedence based on their relative positions, inserting the corresponding directed edges into $C_h^{\mathrm{raw}}$ or $C_v^{\mathrm{raw}}$. 
Then $\textsc{Closure}(\cdot)$ performs a transitive closure to obtain the complete spatial reachability graph, and $\textsc{Reduce}(\cdot)$ applies a transitive reduction to prune all redundant relations, resulting in the minimal but equivalent backbone of the rTCG~\cite{Lin_TVLSI04_TCG}. 

\input{algo/Algo_TCG}

Each node feature $\mathbf{x}_v$ encodes both geometric and material properties of die~$v$, including its normalized width, height, centroid location, the bare-die CTE, the effective CTE of the die–underfill composite, obtained through the thickness-weighted mixing function $f_{\text{eff}}$, and the CTE mismatch
between the composite and the substrate. 
This provides a compact representation of multi-level thermo-mechanical coupling at the die scale.
Meanwhile, each edge feature $\mathbf{e}_{uv}$ captures the relative spatial relationship and material contrast between neighboring dies~$(u,v)$, including their horizontal/vertical adjacency type, normalized relative displacements, gap and overlap metrics, and the effective CTE difference, which jointly characterize local geometry and thermal interactions.

\subsection{GCN-based \textit{WarPGNN} Architecture for Parametric Thermal Warpage Prediction}
\label{sec:gcn_warpGNN}

In this section, we present \textbf{\textit{WarPGNN}}, a parametric framework designed for zero-shot warpage prediction under multi-physics coupling. 
The framework takes the reduced TCG $rTCG(V,E)$ with floorplan features as input, and outputs both a deformation map and warpage value defined in Eq.~\eqref{eq:warpage_def}.
\textbf{\textit{WarPGNN}} model consists of two main components, an $L$-layer GCN-based encoder that extracts hierarchical graph embeddings, and a U-Net–inspired decoder that reconstructs the pixel-level deformation field from graph features. Fig.~\ref{fig:GNN_arch} illustrates the node update of a single GCN layer, where all MLPs are learnable.

For each edge feature $\mathbf{e}_{uv}$ connecting nodes $(u, v)\in \mathbf{E}$ in the $l$-th layer, we calculate the edge weight $g_{uv}$ using a 2-layer MLP expressed by Eq.~\eqref{eq:scaler_gate},
\begin{equation}
\small
    g_{uv}^{(l)} = \sigma\Big((\mathbf{w}_2^{(l)})^\top \,\text{ReLU}(\mathbf{W}_1^{(l)}\cdot \mathbf{e}_{uv} + \mathbf{b}_1^{(l)}) + \mathbf{b}_2^{(l)} \Big)
\quad \in (0,1)
\label{eq:scaler_gate}
\end{equation}
where $\sigma$ denotes the activation function, $(\mathbf{w}_2^{(l)})^\top, \mathbf{W}_1^{(l)}, \mathbf{b}_1^{(l)}, \mathbf{b}_2^{(l)}$ are learnable parameters, ReLU$(\cdot)$ is an activation function.
\vspace{-8pt}
\input{figtex/GNN_arch}

The message $\mathbf{m}_v^{(l)}$ passing in the $l$-th layer can be calculated by Eq.~\eqref{eq:message-passing} with a trainable layer-specific parameter $\mathbf{W}^{(l)}$,

\begin{equation}
\small
\mathbf{m}_v^{(l)} = \sum_{u \in \mathcal{N}(v)} g_{uv}^{(l)} \cdot \mathbf{W}^{(l)} \mathbf{h}_u^{(l)}
\label{eq:message-passing}
\end{equation}
where $\mathcal{N}(v)$ are the neighboring set of node $v$ and $\mathbf{h}_u^{(l)}$ denotes the embedding of node $u\in \mathcal{N}(v)$ obtained from the previous layer.
Here we define $\mathbf{h}_v^{(1)}=\mathbf{x}_v$, $v\in V$.
To stabilize training and incorporate global context into each GCN layer, we apply Eq.~\eqref{eq:aggregation} to jointly aggregate localized and global representations through skip connections.
\begin{equation}
\small
    \tilde{\mathbf{m}}_v^{(l)} = (1+\gamma_v^{(l)}) \odot \mathbf{m}_v^{(l)} + \beta_v^{(l)}
    \label{eq:aggregation}
\end{equation}
here $\odot$ denotes channel-wise multiplication, terms $\gamma_v^{(l)}, \beta_v^{(l)}$ are learnable parameters given by Eq.~\eqref{eq:gamma_beta},
\begin{equation}
\small
    [\gamma_v^{(l)}, \beta_v^{(l)}] = \mathbf{W}_2\, \phi(\mathbf{W}_1 \mathbf{s}_v + \mathbf{b}_1) + \mathbf{b}_2
    \label{eq:gamma_beta}
\end{equation}
in which $\mathbf{s}_v$ denotes the broadcasted global packaging context such as thermal load and substrate properties serving as skip information, and $\phi(\cdot)$ represents non-linear activation function.
This operation represents a Multi-layer Perceptron (MLP) network for Feature-wise modulation~\cite{perez2018film}.
Furthermore, we apply a residual connection and layer normalization~\cite{ba2016layernormalization} on the derived message $\tilde{\mathbf{m}}_v^{(l)}$, as shown in Eq.~\eqref{eq:layer_output},
\begin{equation}
\small
    \mathbf{h}_v^{(l+1)} = \text{LayerNorm}(\text{ReLU}(\tilde{\mathbf{m}}_v^{(l)} + \mathbf{r}_v^{(l)})).
    \label{eq:layer_output}
\end{equation}
with the introduced residual term ${\small \mathbf{r}_v^{(l)}=\mathbf{W}^{(l)}_{r}\mathbf{h}_v^{(l)}}$ and $\mathbf{W}^{(l)}_{r}$ as a layer-specific trainable parameter.

After the final $L$-th layer of GCN-based encoder, the node embedding $\mathbf{h}_v^{(L+1)}$ of each node $v\in V$ encodes both its local attributes and its global structural context.
To enable convolutional decoding, these die-level embeddings must be translated into a spatial representation.
Therefore, we construct a fixed-resolution H $\times$ W grid covering the entire package and project each die embedding onto the grid region corresponding to its normalized physical positions.
For each point (x,y) on the H$\times $W grid, we locate the die whose normalized footprint covers that position.
Each grid point falling inside die $v\in V$ is assigned the node embedding $\mathbf{h}_v^{(L+1)}$, while grid points outside all dies are set to zero to indicate substrate.
This operation yields a spatial feature tensor $\mathbf{F} \in \mathbb{R}^{d_h \times H \times W}$ that is piecewise constant over each die region and preserves the floorplan geometry, where $d_h$ denotes the dimensionality of the node embeddings $\mathbf{h}_v^{(L+1)}$.

The feature field $\mathbf{F}$ is then passed to a U-Net based decoder $\mathcal{D}_\theta(\cdot)$, which employs a lightweight multi-scale convolutional architecture with symmetric skip connections, enabling effective capture of both local and global warpage patterns, to reconstruct the deformation $\hat{{w}}^{H\times W}$ as shown in Eq.~\eqref{eq:decoder},
\begin{equation}
\small
\hat{{w}}^{H\times W} =
\mathcal{D}_\theta(\mathbf{F})
\label{eq:decoder}
\end{equation}

\subsection{Physics-aware Loss and Graph Representation Enhancement for Learning Improvement}

We observe that, owing to the PDE-governed nature of the warpage formulation, the deformation values exhibit a highly non-uniform numerical distribution, resembling the hyperbolic-shape solution described in~\cite{Tsai_TCPMT20}.
As shown in Fig.~\ref{fig:WarpageStats}, most of normalized deformation values fall within the interval [-2, 2], with a small number of outliers form a distinct long-tailed behavior that hinder the model’s ability to learn extreme cases effectively.
Fig.~\ref{fig:error_map} further illustrates this effect.
When trained with a vanilla MSE loss, the GNN exhibits large prediction errors in the top-left region of the package.
These high-error zones coincide with the outlier deformation values in the long-tailed distribution, confirming the model’s difficulty in capturing extreme cases, which is precisely the most critical regime for warpage analysis.

\input{figtex/Error_map}

To mitigate the data-scarcity issue and improve learning performance, we further developed a Physics-informed loss function by introducing penalty terms to confine the optimization space of \textit{\textbf{WarPGNN}} model. 
We employ two additional terms in the loss function based on our observations in warpage data statistics to perform training.

\blackcircled{1}\textbf{Tail-aware reweighting strategy}. 
Because a vanilla MSE loss indiscriminately averages over all grid cells, the model tends to favor the dominant low-magnitude regions and underfit the sparse yet critical tail cases. 
To alleviate this imbalance, we introduce a tail-selection mask $m = \mathbf{1}\cdot\big(|w| > \sigma_{\mathrm{t}}\big)$ based on a prescribed threshold $\sigma_{\mathrm{t}}$ to screen high-magnitude deformation regions, where $w$ is the ground-truth value.
The resulting penalty term is defined as in Eq.~\eqref{eq:tail_loss},
\begin{equation}
\small
    \mathcal{L}_{\mathrm{tail}} = \frac{1}{N} \sum_i^N m_i \|\hat{w}_i - w_i\|^2
    \label{eq:tail_loss}
\end{equation}

\blackcircled{2}\textbf{Physics-aware gradient matching}. To enhance the model’s ability to capture local curvature and high-gradient variations in the deformation field, we employ the Sobel operator~\cite{Gonzalez_DIP} to estimate spatial gradients on the discretized deformation map.
Gradients $\nabla w=[\nabla_x w,\nabla_y w]^T$ are computed via $\nabla_x w = S_x * w,\ \nabla_y w = S_y * w$, where $S_x$ and $S_y$ denote the horizontal and vertical Sobel kernels and $*$ is convolution.
The predicted and ground-truth gradients are enforced to match through the following penalty term shown in Eq.~\eqref{eq:grad_loss},
\begin{equation}
\small
    \small \mathcal{L}_{\mathrm{grad}}= \left\| \nabla_x \hat{w} - \nabla_x w \right\|_2^2+ \left\| \nabla_y \hat{w} - \nabla_y w \right\|_2^2
    \label{eq:grad_loss}
\end{equation}

The enhanced PI-loss function $\mathcal{L}_{\text{PI}}$ is defined in Eq.~\eqref{eq:weighted_loss},
\begin{equation}
\small
    \mathcal{L}_{\text{PI}}=\mathcal{L}_{\text{MSE}}+\alpha\cdot\mathcal{L}_{\text{tail}}+\lambda\cdot \mathcal{L}_{\text{grad}}
    \label{eq:weighted_loss}
\end{equation}
where $\mathcal{L}_{\text{MSE}}$ is the vanilla MSE loss, $\alpha$ and $\lambda$ denote coefficients that balance the learning performance between MSE and PI penalty terms. 

To enhance the expressive capability of the graph encoder and better capture
the coupling between multi-physics and geometry of multi-die floorplans, we adopt a Graph Isomorphism Network (GIN)~\cite{Xu_ICLR19_GIN}-based
encoder that replaces the message-passing in Eq.~\eqref{eq:message-passing}
with an injective neighborhood aggregation scheme. 
The revised encoder leverages the GIN formulation, which alleviates the over-smoothness in GCN prediction tasks and thus becomes suitable for \textbf{\textit{WarPGNN}}.

Given the input graph $rTCG(V,E)$ from Algo.~\ref{alg:rTCG_short} and initial node features $\mathbf{h}_v^{(1)}$ for each node $v\in V$, a GIN encoder with $L$ layers generates a hierarchy of node embeddings $\mathbf{h}_v^{(2)},\ldots,\mathbf{h}_v^{(L+1)}$, each enriching the local representation with broader structural context.

For a node $v\in V$ in the $l$-th layer, information is aggregated from its neighbors $\mathcal{N}(v)$ using a learnable scalar $\epsilon^{(l)}$ as shown in Eq.~\eqref{eq:gin-agg},
\begin{equation}
\small
    \mathbf{a}_v^{(l)}=(1+\epsilon^{(l)})\,\mathbf{h}_v^{(l)}+\sum_{u\in\mathcal{N}(v)}\mathbf{h}_u^{(l)},\ l\in[1,L]
    \label{eq:gin-agg}
\end{equation}

The aggregated feature $\mathbf{a}_v^{(l)}$ is then transformed through a MLP with residual connection to ensure injective mapping and enhance the learning performance of deeper layers,
\begin{equation}
\small
\tilde{\mathbf{h}}_v^{(l)}=
\mathrm{ReLU}\!\left(W_{2}^{(l)}\mathrm{ReLU}(W_{1}^{(l)}\mathbf{a}_v^{(l)})\right)
+
W_{\mathrm{res}}^{(l)} \mathbf{a}_v^{(l)},
\label{eq:gin-mlp}
\end{equation}
where $W_{\mathrm{res}}^{(l)} = I$ if the input and output dimensions match.

To stabilize the activation statistics across graphs of varying sizes, the intermediate representation is normalized by LayerNorm and then activated by ReLU,
\vspace{-2pt}
\begin{equation}
    \small
\mathbf{h}_v^{(l+1)} =\mathrm{ReLU}\!\left(\mathrm{LayerNorm}\!\left(\tilde{\mathbf{h}}_v^{(l)}\right)\right)
\label{eq:gin-layer-out}
\end{equation}

The final-layer node embeddings $\{\mathbf{h}_v^{(L+1)}\}_{v\in V}$ will then be processed as in Sec.~\ref{sec:gcn_warpGNN}.
Our GIN-based \textbf{\textit{WarPGNN}} inherits the efficiency from GCN-based model and has improved learning ability by the revised data aggregation.

\subsection{Overall Training Framework of \textit{\textbf{WarPGNN}}}

\input{figtex/Framwork}
The overall training workflow is illustrated in Fig.\ref{fig:overall_famework}.
For each multi-die floorplan, ground-truth warpage are generated using 3-D FEM solvers (e.g., COMSOL). 
In parallel, the same floorplans are converted into their rTCG format following in Algo.~\ref{alg:rTCG_short}, which serve as inputs to our \textbf{\textit{WarPGNN}} model.
Prior to full-scale training, a small subset of training data is used to identify the optimal hyperparameters \{$\alpha^*, \lambda^*$\} that minimize the converged PI-loss and balance the contributions from data and physics-consistency.
In this study we sweep $\alpha, \lambda$ from 0 to 0.1 and select \{$\alpha^*, \lambda^*$\}=\{0.05, 0.02\}.
Once these hyperparameters are fixed, the model is trained end-to-end using the PI-loss to accurately learn the deformation across diverse multi-die configurations.

%% file: algo/Algo_TCG.tex
\vspace{-7pt}
\begin{algorithm}[htp]
\caption{rTCG Construction for Multi-Die Floorplan}
\small
\label{alg:rTCG_short}
\begin{algorithmic}[1]
\Require Multi-die floorplan $\mathcal{D} = \{d_i\}_{i=1}^N$ with die sizes $(w_i,h_i)$, CTE $\alpha_i$ and bottom-left corner position $(x_{o,i}, y_{o,i})$; 
         material parameters $(E_{\text{die}}, t_{\text{die}})$, $(E_{\text{fill}}, \alpha_{\text{fill}}, t_{\text{fill}})$,
         $(E_{\text{sub}}, \alpha_{\text{sub}}, t_{\text{sub}})$; package size ($W_{\text{pkg}}$, $H_{\text{pkg}}$).
\Ensure Attributed graph $rTCG=(V,E)$ for GNN training.
\State $(C_h^{\text{raw}}, C_v^{\text{raw}}) \leftarrow \textsc{InsertPrecedeEdge}(i,j)$ for all $1 \le i < j \le N$
\State $(C_h^{\text{red}}, C_v^{\text{red}}) \leftarrow \textsc{Reduce}(\textsc{Closure}(C_h^{\text{raw}}, C_v^{\text{raw}}))$
\State $V \leftarrow \{1,\dots,N\}$, $E \leftarrow \{(u,v)\mid (u,v)\in C_h^{\mathrm{red}} \cup C_v^{\mathrm{red}}\}$
\For{$v\in V$}  \Comment{Node feature}
    \State $\tilde{w}_v, \tilde{h}_v, \tilde{x}_{c,v}, \tilde{y}_{c,v}= \frac{w_v}{W_{\mathrm{pkg}}}, \frac{h_v}{H_{\mathrm{pkg}}}, \frac{x_{o,v} + 0.5\cdot w_v}{W_{\mathrm{pkg}}}, \frac{y_{o,v} + 0.5\cdot h_v}{H_{\mathrm{pkg}}}$
    \State $\alpha_v^{\text{eff}},\ \Delta \alpha_v \leftarrow f_{\text{eff}}(E_{\text{die}}, t_{\text{die}}, E_{\text{fill}}, \alpha_{\text{fill}},\ t_{\text{fill}}, \alpha_v),\ \alpha_v^{\text{eff}} - \alpha_{\text{sub}}$
    \State $\mathbf{x}_v \leftarrow \operatorname{concat}(\tilde{w}_v,\tilde{h}_v,\tilde{x}_{c,v},\tilde{y}_{c,v},\alpha_v,\alpha_v^{\text{eff}},\Delta\alpha_v)$
\EndFor
\For{each $(u,v) \in E$}  \Comment{Edge feature}
    \State $\tau_{uv} \leftarrow \text{type}(u, v) \in \{\text{H},\text{V}\}$  
    \State $\Delta x_{uv},\Delta y_{uv},\rho_{uv} = \tilde{x}_{c,v} - \tilde{x}_{c,u},\ \tilde{y}_{c,v} - \tilde{y}_{c,u},\frac{\max\big(0,\ x^{\text{L}}_v - x^{\text{R}}_u\big)}{W_{\text{pkg}}}$
    \State $\boldsymbol{\delta}_{uv},\Delta\alpha_{uv} \leftarrow (\Delta x_{uv}, \Delta y_{uv}, \rho_{uv}),\ \alpha_u^{\text{eff}} - \alpha_v^{\mathrm{eff}}$ 
    \State $\Delta w_{uv}, \Delta h_{uv}\leftarrow (\tilde{w}_u - \tilde{w}_v),\ (\tilde{h}_u-\tilde{h}_v) $
    \State $\mathbf{e}_{uv} \leftarrow \operatorname{concat}(\tau_{uv}, \boldsymbol{\delta}_{uv}, \Delta\alpha_{uv}, \Delta w_{uv}, \Delta h_{uv})$
\EndFor
\State \Return $rTCG (V,E) \leftarrow\{\mathbf{x}_v\}^N,\{\mathbf{e}_{uv}\}_{(u,v)\in E}$
\end{algorithmic}
\end{algorithm}
\vspace{-10pt}

%% file: figtex/GNN_arch.tex
\begin{figure}[h]
    \centering
    \includegraphics[width=0.9\linewidth]{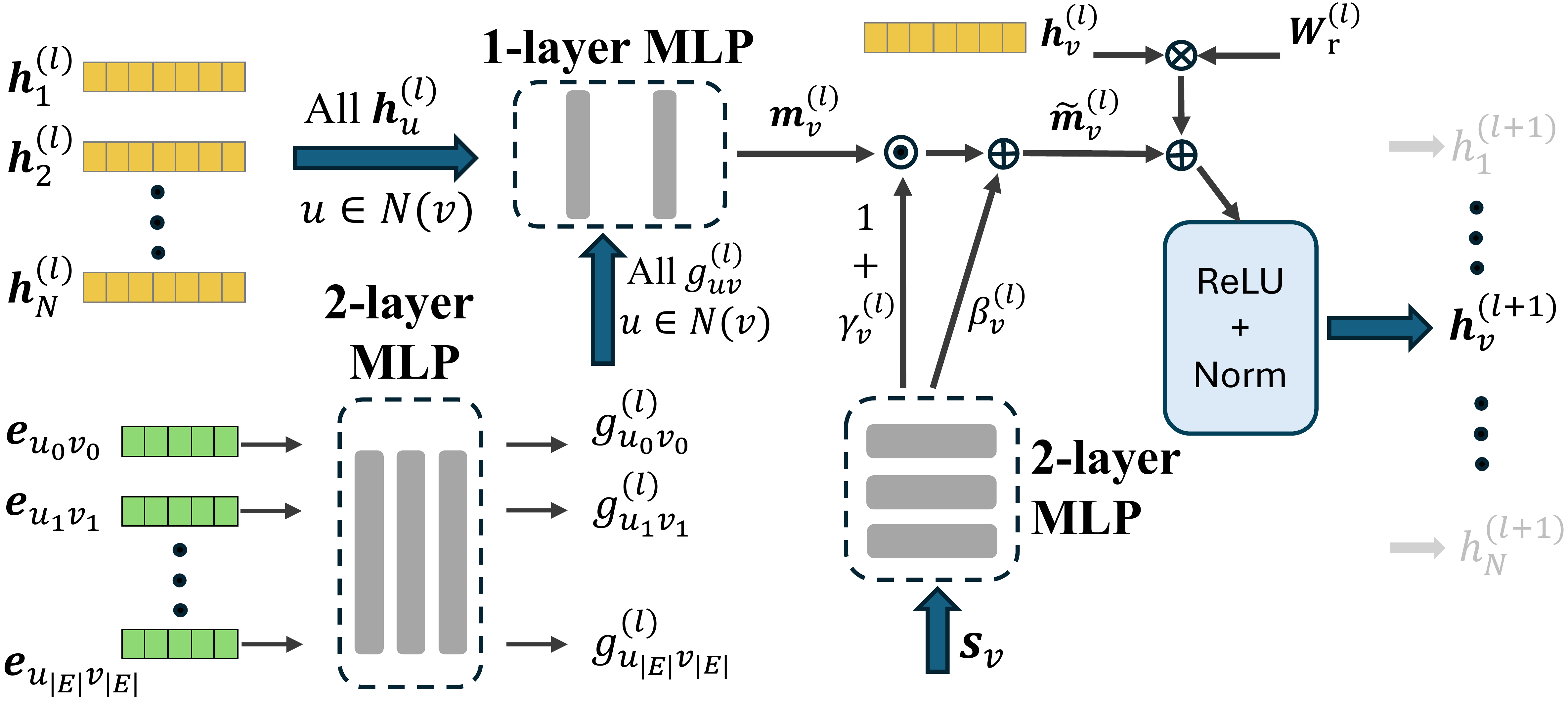}
    \vspace{-10pt}
    \caption{\small Illustration of node embedding update within a single $l$-th layer of our proposed GCN-based encoder.
    $\textbf{h}_i^{(l)}$ are node embeddings.}
    \vspace{-10pt}
    \label{fig:GNN_arch}
\end{figure}

%% file: figtex/Error_map.tex

\begin{figure}[htp]
    \centering
    \vspace{-15pt}
    \hspace{-0.05\columnwidth} 
    \subfloat[]{
        \centering
        \includegraphics[height=3.6cm]{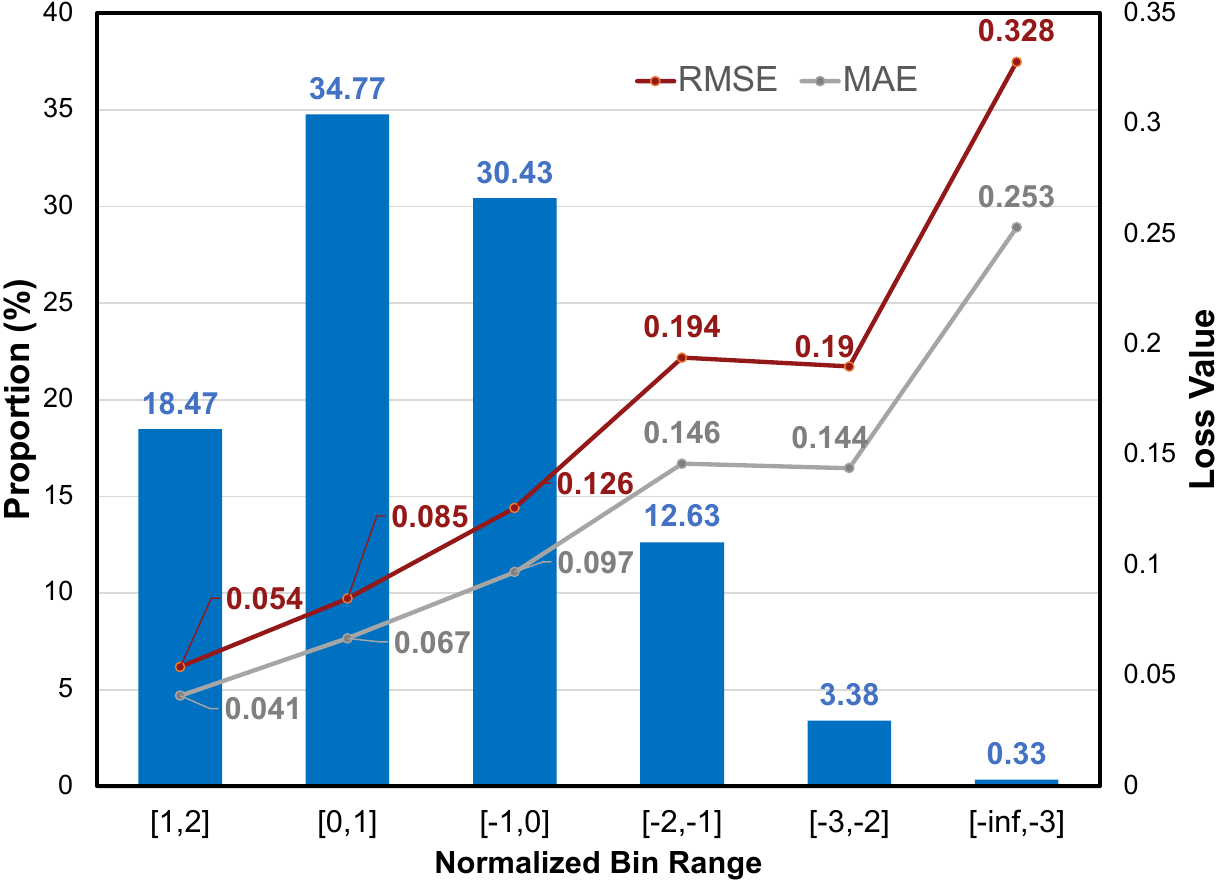}
        \label{fig:WarpageStats}
    }
    \subfloat[]{
        \centering
        \includegraphics[height=3.6cm]{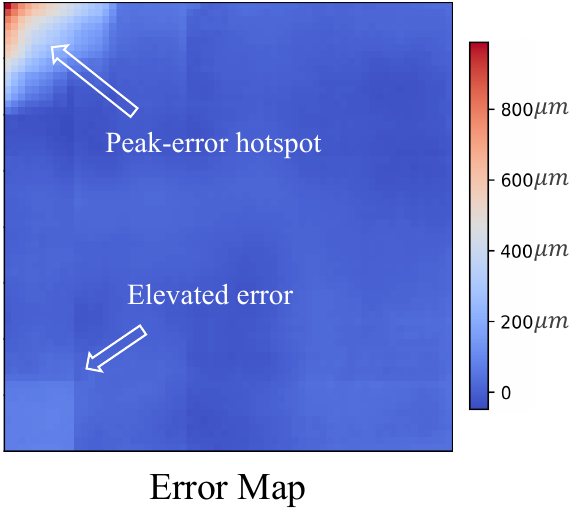}
        \label{fig:error_map}
    }
    \vspace{-10pt}
    \caption{\small (a) Distribution statistics and model performance on normalized ground-truth results. 
    (b) An example of the error map between GCN-based \textit{WarPGNN} prediction and ground-truth deformation map.}
    \vspace{-10pt}
\end{figure}

%% file: figtex/Framwork.tex
\begin{figure*}[!htp]
    \centering
    \includegraphics[width=0.95\linewidth]{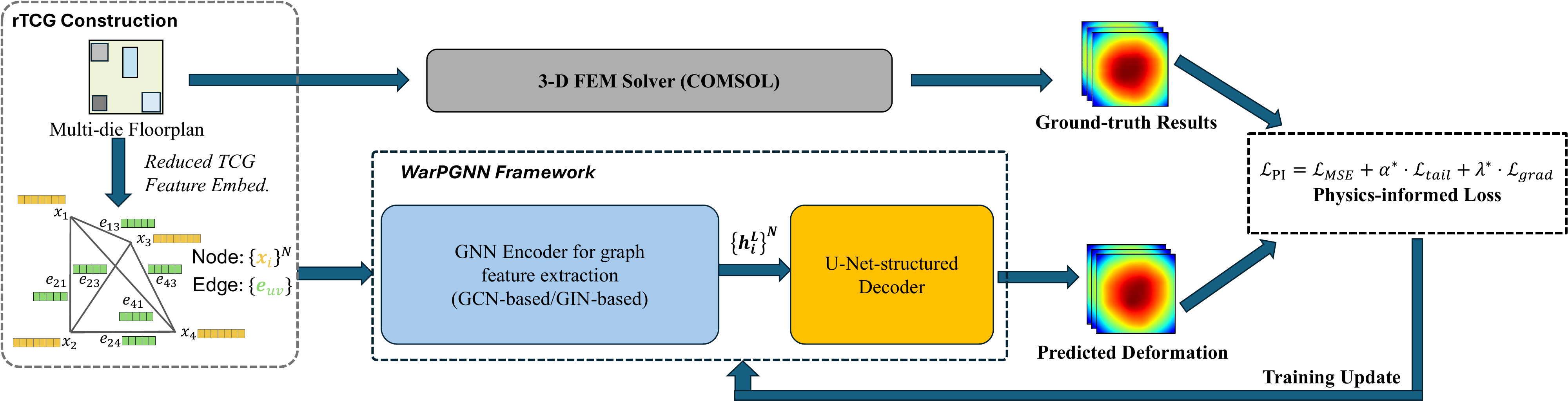}
    \vspace{-10pt}
    \caption{\small The overall framework of our proposed \textit{\textbf{WarPGNN}} for efficient and accurate thermal warpage analysis}
    \label{fig:overall_famework}
    \vspace{-5pt}
\end{figure*}

%% file: doc/4result.tex
\section{Experimental Results}
\label{sec:ExperimentalResults}

\subsection{Experimental Setup}

\textbf{Model}: The proposed \textit{\textbf{WarPGNN}} framework is fully developed using Python/PyTorch and DGL library. 
The training and test procedures are conducted on a Linux server equipped with two Intel 22-core E5-2699 CPUs and an Nvidia TITAN RTX GPU.
Properties of materials and geometry of assembly used in this section are consistent with~\cite{Tsai_TCPMT20, Lo_ICCAD24_EfficientWarpage} and listed in Tab.~\ref{tab:matproper}.
Thermal load $\Delta T$ is fixed at -115K to mimic the drop from fabrication temperature 200$^o$C to peak working temperature 85$^o$C.
Our proposed GCN-based and GIN-based \textbf{\textit{WarPGNN}} have 4 and 3 layers, respectively, and were trained for 53 and 57 epochs.

\textbf{Evaluation Metrics}: We employ normalized root-mean-square-error (RMSE) and mean-absolute-error (MAE) calculated per-sample as performance metrics to evaluate the difference between deformation map prediction and ground-truth results obtained from COMSOL. 
Error of warpage value is evaluated using absolute error.
Furthermore, training and inference time (runtime) is utilized to demonstrate the computational efficiency and speedup of our proposed method.  
\input{tables/MatProper}

\textbf{Data Normalization}:
In our work, input graphs are preprocessed using parameter-wise Min-Max Normalization. Ground-truth deformation maps are normalized using Z-score Standardization~\cite{Bishop:book'06} with dataset-level mean $\mu$ and standard deviation $\sigma$.
After obtaining the prediction maps from \textbf{\textit{WarPGNN}} model, we perform denormalization with $\mu$ and $\sigma$ to derive real deformation map. 

\subsection{Dataset Generation}
To construct training and test datasets, we employ 3-D FEM-based COMSOL~\cite{comsol:heat'2014} to produce accurate thermal warpage simulation serving as ground-truth results for model training following~\cite{Hsu_ICCAD22_TCGFloorplan, Lo_ICCAD24_EfficientWarpage}. 
To diversify spatial and material configurations for training, we employ an assignment algorithm that jointly samples position and CTE for each die.
The chip area is divided into $4\times4$ uniform sub-regions each with $50\times50$~mm$^2$, and one die is randomly placed within each region, with its CTE value uniformly sampled from 3–11~$\mu$K$^{-1}$ to emulate realistic material heterogeneity.
\vspace{-5pt}
\input{tables/TSDataset}

We fix the output deformation map as 64$\times$64 for simplicity.
Our synthesized dataset contains 200 multi-die floorplans each with 40 CTE assignments.
The total 200 floorplans are randomly split into training and test sets with 85\% and 15\%.
To demonstrate the transferability of our proposed \textbf{\textit{WarPGNN}}, we further generated four extended datasets each containing 1200 random designs.
Tab.~\ref{tab:TSDataset} shows the specifications of extended datasets.

\subsection{Ablation Study: Performance of the Proposed GCN-based \textbf{\textit{WarPGNN}} on Seen Datasets}

To evaluate the performance of the proposed GCN-based \textbf{\textit{WarPGNN}} and the effect of the PI-loss, we conduct an ablation study by training the model with and without the PI constraint under identical configurations.
Tab.~\ref{tab:ablation} summarizes the comparison against COMSOL~\cite{comsol:heat'2014}.

\input{tables/ablation_GCN}

For the GCN-based model trained solely with the vanilla MSE loss, the network attains an average RMSE of 74.18~$\mu$m on the test set, corresponding to a 3.46\% error rate over the full deformation range of 2144.10~$\mu$m.
The resulting warpage value prediction exhibits an average deviation of 67.18~$\mu$m, equivalent to a 4.54\% relative error.
In terms of computational efficiency, the proposed GCN-based \textbf{\textit{WarPGNN}} achieves an average inference time of 3.22~ms, offering over 54192$\times$ acceleration compared with COMSOL which requires 174.50 seconds per simulation.
These evaluations validate that the proposed model can serve as a high-fidelity and efficient surrogate for thermal warpage analysis.
Furthermore, by incorporating the proposed PI-loss, the model achieves remarkable gain in accuracy, reducing the average warpage error to 2.33\% and the RMSE to 1.89\%, respectively, while maintaining over 54874$\times$ speedup over COMSOL.

\input{tables/mainresult}

\subsection{Overall Performance of the proposed GIN-based \textbf{\textit{WarPGNN}}}

To demonstrate the accuracy and inference speedup of the proposed method, we apply the GIN-based \textbf{\textit{WarPGNN}}, GCN-based \textbf{\textit{WarPGNN}} trained with the PI-loss to capture deformation of the test set with 1200 samples.
The comparison of our proposed \textbf{\textit{WarPGNN}} warpage simulator with COMSOL~\cite{comsol:heat'2014}, 2-D FEM solver~\cite{Lo_ICCAD24_EfficientWarpage} and DeepONet-based solver~\cite{Lo_ICCAD25_DeepONet} are shown in Tab.~\ref{tab:mainresults}. 
Here we directly adopted the reported metrics of the most similar benchmark from~\cite{Lo_ICCAD24_EfficientWarpage} as the implementation is not publicly available for comparative reference.
However, as our test cases incorporate a more sophisticated \textit{die–underfill–substrate} stack and more on-chip dies, the quoted metrics represent a conservative reference for comparative evaluation.
DeepONet is re-implemented to match the same die–underfill–substrate stack configuration and trained on the same training and test sets without data augmentation for fair numerical comparison.

Results show that our proposed GIN-based \textbf{\textit{WarPGNN}} further enhances prediction fidelity with supreme computational efficiency.
The average warpage error decreases to 35.16~$\mu$m, corresponding to a relative error of 2.21\%, while the average RMSE is reduced to 26.94~$\mu$m, equivalent to 1.26\% error rate.
Furthermore, \textbf{\textit{WarPGNN}} achieves an average inference time of only 1.46~ms, yielding a 119766.64$\times$ speedup over the 3-D FEM COMSOL~\cite{comsol:heat'2014} and a 205.91$\times$ acceleration over the 2-D efficient FEM method~\cite{Lo_ICCAD24_EfficientWarpage}.
Compared with~\cite{Lo_ICCAD25_DeepONet}, our GIN-based method achieves similar prediction accuracy and speedup on inference, while reducing the training time by 70.4\%.

\input{figtex/Main_fig}

Fig.~\ref{fig:main_fig} provides visual comparisons between the predicted and ground-truth deformation maps, where white squares denote die and underfill.
As shown in Fig.~\ref{fig:main_fig}, \textit{Design 1} resembles the same floorplan depicted in Fig.~\ref{fig:error_map}, the GIN-based \textbf{\textit{WarPGNN}} achieves improved prediction accuracy and noticeably alleviates prediction errors in high-deformation areas with data scarcity.

Our results indicate that the proposed model can predict deformation and warpage value with high fidelity and superior efficiency across diverse multi-die floorplans featuring various material properties in advanced packaging systems.

\subsection{Transferability of the proposed \textbf{\textit{WarPGNN}}} 

Tab.~\ref{tab:transfer_results} summarizes the transferability performance of the proposed GIN-based \textbf{\textit{WarPGNN}} model in comparison with COMSOL on four extended datasets \ding{172} to \ding{175}.
The numerical ranges of the average deformation field obtained from four extended datasets are 2083.18$\mu$m, 2614.64$\mu$m, 2093.09$\mu$m and 2182.96$\mu$m, respectively.
Across different cases, our model achieves an average RMSE of 46.73$\mu$m, corresponding to a 2.32\% error rate in case \ding{172}, and a maximum RMSE of 80.56$\mu$m translating to 3.69\% error in cases \ding{173}–\ding{175}.
\input{tables/Transfer_result}

The results demonstrate that the proposed model remains robust under parameter shifts and retains transferability.
Although the accuracy on extended cases degrades by up to 2.94$\times$ relative to the original test set, it is still sufficient for early-stage warpage-aware floorplan optimization~\cite{Hsu_ICCAD22_TCGFloorplan}.
Even under such distribution shifts, the achieved error levels are still well within the range required for reliable warpage analysis and utilizable for practical warpage-aware floorplan optimization.

Fig.~\ref{fig:transfer_result} shows examples of the predicted deformation obtained from \textbf{\textit{WarPGNN}} on the extended datasets, where the unit is $\mu$m.
Results from our model align closely with the ground-truth values, suggesting the transferability of our proposed method under extended cases.
\vspace{-5pt}
\input{figtex/transfer_results}

%% file: tables/MatProper.tex
\begin{table}[htp]
\vspace{-7pt}
\centering
\caption{\small Material and geometric parameters of the simulated assembly for warpage analysis}
\vspace{-10pt}
\resizebox{0.88\linewidth}{!}{
\begin{tabular}{c|ccc|cc}
\hline\hline
\multirow{2}{*}{Object} & \multicolumn{3}{c|}{Material Property} & \multicolumn{2}{c}{Geometry}\\ \cline{2-6} 
                        & CTE ($\mu$K$^{-1}$)        & $\nu$    & E (Gpa)    & Thickness (mm) & Length (mm) \\ \hline\hline
Die                     & 3$\sim$11  & 0.28  & 169  & 0.875  & 25  \\
Underfill               & 29         & 0.32  & 7.6  & 0.06   &  25 \\
Substrate               & 13.2       & 0.42  & 20   & 1.81   &  200 \\ \hline\hline
\end{tabular}
}
\label{tab:matproper}
\vspace{-8pt}
\end{table}

%% file: tables/TSDataset.tex
\begin{table}[htp]
\centering
\caption{\small Details of the four types of extended datasets with corresponding change of parameter}
\vspace{-10pt}
\resizebox{0.7\linewidth}{!}{
\begin{tabular}{c|c|c}
\hline\hline
Case & Change of Parameters & Parameter Space \\ \hline\hline
\ding{172}    & CTE of dies: $\alpha_{\text{die}}$         & $[3, 11] \rightarrow [2, 13] \, \mu\text{K}^{-1}$             \\ 
\ding{173}    & Size of dies: $L_{\text{die}}$, $W_{\text{die}}$       & $25 \rightarrow [20, 45] \, \text{mm}$           \\
\ding{174}    & Size of chip: $L_{\text{chip}}$=$W_{\text{chip}}$        & $200 \rightarrow [160, 240] \, \text{mm}$            \\
\ding{175}    & Number of Die        & $16 \rightarrow \{15, 20\}$            \\ \hline\hline
\end{tabular}
}
\label{tab:TSDataset}
\vspace{-10pt}
\end{table}

%% file: tables/ablation_GCN.tex
\begin{table}[htp]
\vspace{-5pt}
\centering
\caption{\small Performance comparison between COMSOL and the proposed GCN-based \textbf{\textit{WarPGNN}} model trained with different loss.}
\vspace{-10pt}
\resizebox{0.92\linewidth}{!}{
\begin{tabular}{c|cccc}
\hline\hline
Method                                 & Warpage Error & Avg. RMSE & Runtime (s) & Speedup \\ \hline\hline
COMSOL~\cite{comsol:heat'2014}       &   -         &  -    &    174.50             &   1$\times$       \\ \hline
\begin{tabular}[c]{@{}c@{}}GCN-based \\\textit{\textbf{WarPGNN} w/o PI-loss}\end{tabular} & 4.54\%           &  3.46\%    &              3.22$\times$10$^{-3}$    &    54192.55$\times$      \\ \hline
\begin{tabular}[c]{@{}c@{}}GCN-based \\\textit{\textbf{WarPGNN} w/ PI-loss}\end{tabular} &  2.33\%          & 1.89\%     &   3.18$\times$10$^{-3}$             & 54874.21$\times$         \\ \hline\hline
\end{tabular}
}
\label{tab:ablation}
\vspace{-10pt}
\end{table}

%% file: tables/mainresult.tex
\renewcommand{\arraystretch}{1.2}

\begin{table*}[htp]
\centering

\caption{\small Comparisons of the performance between our proposed GCN- and GIN-based \textbf{\textit{WarPGNN}} and 3-D FEM solver COMSOL~\cite{comsol:heat'2014}, 2-D efficient FEM solver~\cite{Lo_ICCAD24_EfficientWarpage} and DeepONet-based solver~\cite{Lo_ICCAD25_DeepONet}. 
Notation $^{\dagger}$ means data directly quoted from~\cite{Lo_ICCAD24_EfficientWarpage} for reference.}
\vspace{-10pt}
\resizebox{0.96\textwidth}{!}{
\begin{tabular}{c|ccccc|ccc}
\hline\hline
\textbf{Method}                                                      & Avg. Warpage Error~$\downarrow$ & Max RMSE~$\downarrow$ & Min RMSE ~$\downarrow$ & Avg. NRMSE~$\downarrow$ & Avg. NMAE~$\downarrow$ & Training (min)~$\downarrow$  & Inf. Runtime (s)~$\downarrow$ & Speedup~$\uparrow$                                      \\ \hline\hline
3-D FEM COMSOL~\cite{comsol:heat'2014}                                                      & -            &    -      &  -         &   -             & -   &   -  & 174.50             &                   1$\times$                            \\ \hline
2-D FEM Solver~\cite{Lo_ICCAD24_EfficientWarpage} &  $\geq$1\%$^{\dagger}$        &  N.A.        &   N.A.       &     N.A.      &   N.A.    &  -  &  $\geq$0.30$^{\dagger}$            & 581.7\begin{tabular}[c]{@{}c@{}}$\times$ to~\cite{comsol:heat'2014}\end{tabular} \\ \hline\hline
DeepONet-based~\cite{Lo_ICCAD25_DeepONet}  &   3.06\%           & 99.09$\mu$m         &  62.26$\mu$m       &       6.12\%      & 5.82\%      &  110.78      &  $1.29\times10^{-3}$  & \begin{tabular}[c]{@{}c@{}}135271.32$\times$ to~\cite{comsol:heat'2014}\\ 232.56$\times$ to \cite{Lo_ICCAD24_EfficientWarpage}\end{tabular} \\ \hline
\hline
Our GCN-based \textbf{\textit{WarPGNN}}  &   2.33\%           & 106.84$\mu$m         &  15.58$\mu$m       &       1.89\%      & 1.30\%    &  47.29 &    $3.18\times10^{-3}$             & \begin{tabular}[c]{@{}c@{}}54874.21$\times$ to~\cite{comsol:heat'2014}\\ 94.34$\times$ to \cite{Lo_ICCAD24_EfficientWarpage}\end{tabular} \\ \hline
Our GIN-based \textbf{\textit{WarPGNN}}  & 2.21\%           & 69.02$\mu$m         &  8.64$\mu$m        &  1.26\%         &  0.97\%       & 32.82  &   $1.46\times10^{-3}$            & \begin{tabular}[c]{@{}c@{}}119766.64$\times$ to~\cite{comsol:heat'2014}\\ 205.91$\times$ to \cite{Lo_ICCAD24_EfficientWarpage}\end{tabular} \\ \hline\hline
\end{tabular}
}
\label{tab:mainresults}
\vspace{-10pt}

\end{table*}

%% file: figtex/Main_fig.tex
\begin{figure}[htp]
\vspace{-3pt}
    \centering
    \includegraphics[width=0.95\linewidth]{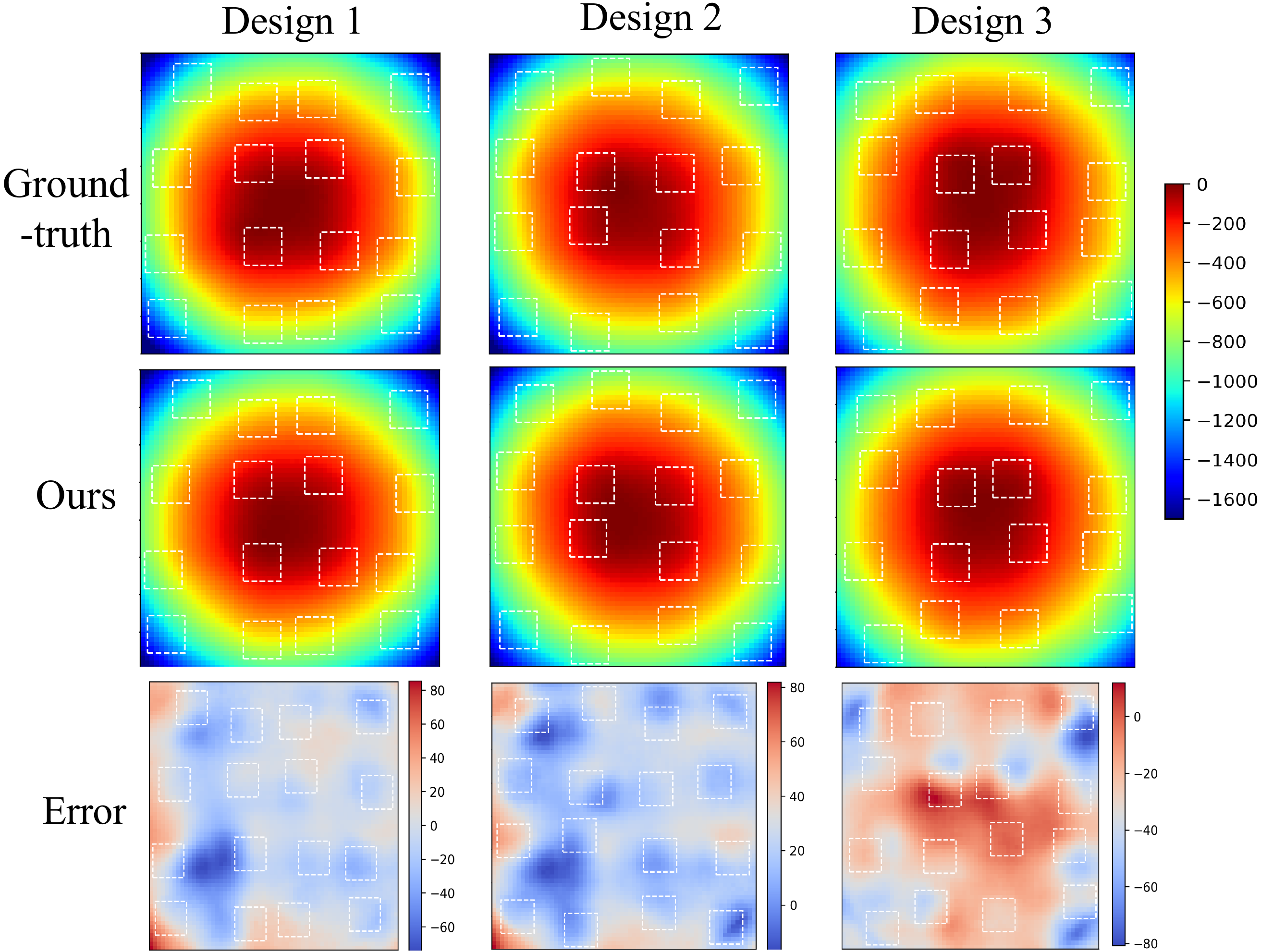}
    \vspace{-10pt}
    \caption{\small Comparison of the predicted deformation map obtained from our GIN-based \textbf{\textit{WarPGNN}} and COMSOL.
    All values are in $\mu$m.}
    \label{fig:main_fig}
    \vspace{-7pt}
\end{figure}

%% file: tables/Transfer_result.tex

\begin{table}[htp]
\centering
\caption{\small Results of the proposed GIN-based \textbf{\textit{WarPGNN}} on extended datasets \ding{172}$\sim$\ding{175} }
\vspace{-12pt}
\resizebox{0.72\linewidth}{!}{
\begin{tabular}{c|cccc}
\hline\hline
Case & \multicolumn{1}{c}{Max RMSE}  & \multicolumn{1}{c}{Min RMSE} & \multicolumn{1}{c}{Avg. NRMSE} & \multicolumn{1}{c}{Avg. NMAE} \\ \hline\hline
\ding{172}     &  104.17$\mu$m                          &  14.27$\mu$m                      & 2.32\%                            &  1.78\%                         \\ \hline
\ding{173}    &   262.33$\mu$m                          &   18.34$\mu$m                      &  2.62\%                             &  1.86\%                        \\
\ding{174}    &   154.18$\mu$m                           &   20.83$\mu$m                          & 3.01\%       & 2.31\%                  \\ 
\ding{175}    &   354.87$\mu$m                           &   20.44$\mu$m                     &  3.69\%      &  2.68\%                  \\ \hline\hline
\end{tabular}
}
\label{tab:transfer_results}
\vspace{-13pt}
\end{table}

%% file: figtex/transfer_results.tex
\begin{figure}[htp]
    \centering
    \includegraphics[width=0.96\linewidth]{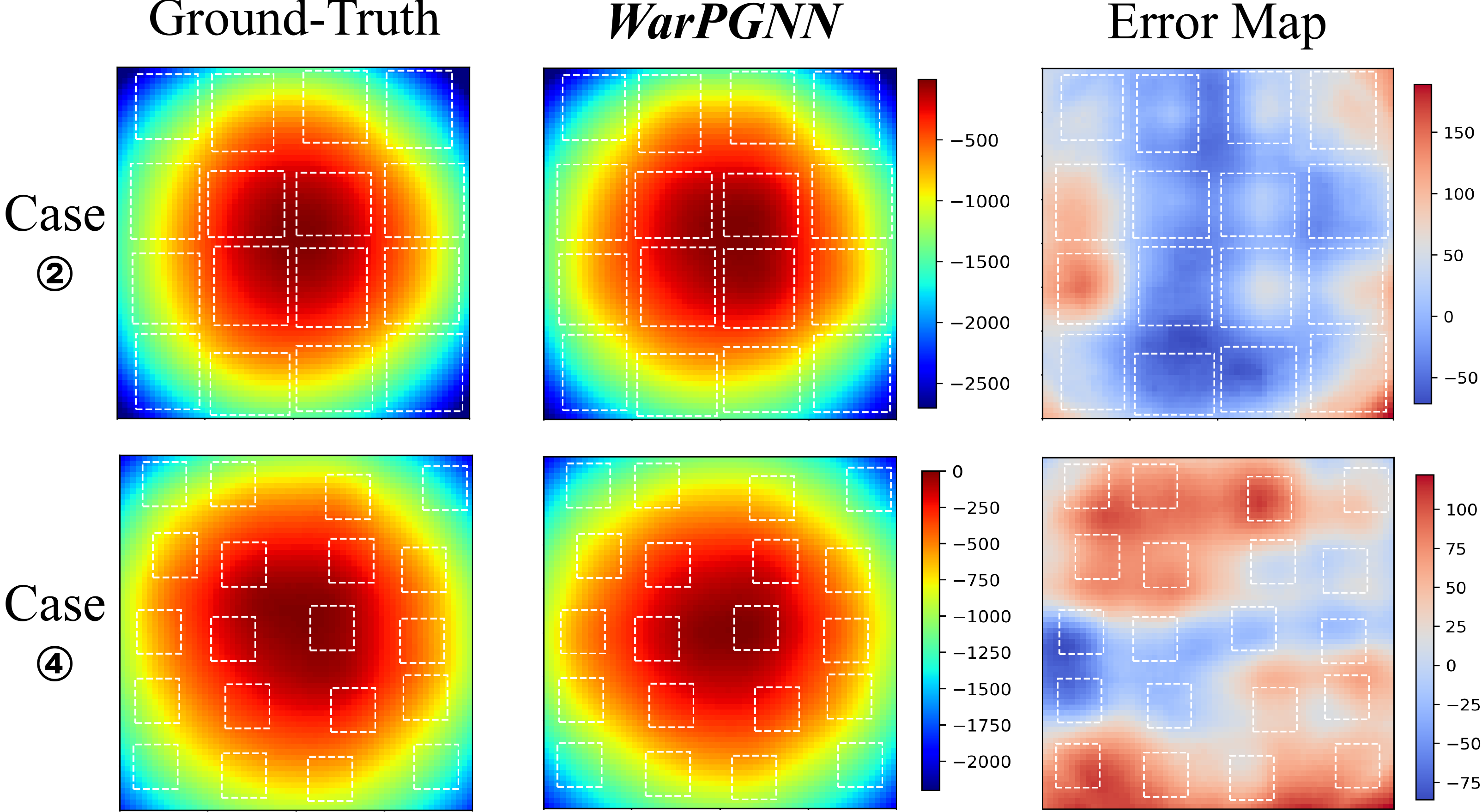}
    \vspace{-10pt}
    \caption{\small Results of our \textit{\textbf{WarPGNN}} framework on extended datasets.}
    \label{fig:transfer_result}
    \vspace{-10pt}
\end{figure}

%% file: doc/5conclu.tex
\section{Conclusion}
\label{sec:Conclusion}
In this paper, we present a novel machine learning-based parametric framework, termed {\it \textbf{WarPGNN}}, to efficiently simulate thermal warpage effects.
In our framework, multi-die floorplans are encoded using rTCG and processed through a GCN-based encoder and U-Net decoder, enabling efficient graph-to-field learning.
Moreover, a Physics-aware loss and GIN encoder that leverages enhanced data aggregation between dies are developed for improving learning performance on both difficult long-tailed cases and graph representation of general die coupling.  
Numerical results show that our proposed GIN-based \textbf{\textit{WarPGNN}} achieves more than 205.91$\times$ and 119767$\times$ runtime speedup compared to 2-D FEM solver and COMSOL, with only 1.26\% full-map normalized RMSE and 2.21\% warpage value error.
Compared with DeepONet, \textbf{\textit{WarPGNN}} improves training efficiency by reducing training time by 70.4\% while maintaining comparable accuracy and inference time.
Furthermore, \textbf{\textit{WarPGNN}} demonstrates transferability on four extended datasets, with at most 3.69\% normalized RMSE and similar inference speed.